\documentstyle[12pt, fleqn]{article}
\begin{document}

\author{\bf {Shahin S. Agaev}\\ \it High Energy Physics Lab., Baku State University\\
\it Z.Khalilov st.23, 370148 Baku, Azerbaijan\thanks{E-mail address:azhep@lan.ab.az}}
\title{\bf{IR RENORMALON EFFECTS AND LIGHT MESONS' ELM FORM FACTOR
}}
\date{}
\maketitle
\begin{abstract}
The pion and kaon electromagnetic form factors $F_{M}(Q^2)$ are calculated
at the leading order of pQCD using the running coupling constant method.
In calculations a dependence of the mesons distribution amplitudes on the hard scale
$Q^2$ is taken into account. The Borel transform and resummed expression for
$F_M(Q^2)$ are found.
\end{abstract}
\newpage

{\bf 1.} Investigation of the infrared (ir) renormalon effects in various 
inclusive and exclusive processes is one of the important and interesting 
problems in the perturbative QCD (pQCD) [1]. It is known that ir renormalons
are responsible for factorial growth of coefficients in perturbative series
for the physical quantities. But these divergent series can be resummed by
means of the Borel transformation [2] and the principal value prescription [3]
and effects of ir renormalons can be taken into account by scale-setting procedure 
$\alpha_{S}(Q^2) \rightarrow \alpha_{S}(exp(f(Q^2))Q^2)$ at the one-loop order
results. Technically, all-order resummation of ir renormalons corresponds to
the calculation of the one-loop Feynman diagrams with the running coupling constant
$\alpha_{S}(-k^{2})$ at the vertices or, alternatively, to calculation of the
same diagrams with non-zero gluon mass. Both these approaches are generalization
of the Brodsky, Lepage and Mackenzie (BLM) scale-setting method [4] and amount
to absorbing certain vacuum polarization corrections appearing at higher-order
calculations into the one-loop QCD coupling constant. Studies of ir renormalon
problems have opened also new prospects for evaluation of power suppressed
(higher twist) corrections to processes' characteristics [5].

Unlike inclusive processes exclusive ones have additional source of ir renormalon
contributions [6-9]. Thus, integration over longitudinal fractional momenta of
hadron constituents in the expression of the electromagnetic (elm) form factor [6-8]
or, in the amplitude of exclusive subprocesses [9] generates ir renormalon
corrections.

In this work we calculate the pion and kaon elm form factors using the running
coupling constant method and by taking into account dependence of the mesons
distribution amplitudes on the hard scale $Q^2$.

{\bf 2.} In the context of pQCD the meson elm form factor has the form,
\begin{equation}
F_M(Q^2)=\int_0^1 \int_0^1 dxdy\phi _M^{*}(y,Q^2)T_H(x,y;Q^2,\alpha _S(\hat{Q}^2))
\phi _M(x,Q^2),\label{1} 
\end{equation}
where $Q^2=-q^2$ is the square of the virtual photon's four-momentum. In Eq.(1)
$\phi_M(x,Q^2)$ is the meson distribution amplitude, containing all non-perturbative 
hadronic binding effects, whereas $T_H(x,y;Q^2,\alpha_{S}(\hat{Q}^2))$
is the hard-scattering amplitude of the subprocess $q \overline{q}'+\gamma^{*}
\rightarrow q\overline{q}'$ and can be found using pQCD. At the leading order
$T_H$ is given by the following expression 
\begin{equation}
T_H(x,y;Q^2,\alpha _S(\hat{Q}^2))=\frac{16\pi C_F}{Q^2}\left[ \frac 23 
\frac{\alpha _S(Q^2(1-x)(1-y))}{(1-x)(1-y)}+\frac 13\frac{\alpha _S(Q^2xy)}{
xy}\right]. \label{2}
\end{equation}

In Eq.(2) $\hat{Q}^2$ is taken as the square of the momentum tranfer of the
exchanged hard gluon in corresponding Feynman diagrams, $C_F=4/3$ is the color
factor.

One of the important ingredients of our study is the choice of the meson
distribution amplitude $\phi _M(x,Q^2)$. In the framework of pQCD it is
possible to predict the dependence of $\phi _M(x,Q^2)$ on $Q^2$ using
evolution equation, but not its shape (its dependence on $x$). An
information about the shape of the distribution amplitude $\phi _M(x,Q^2)$
can be obtained by means of the QCD sum rules method or other
non-perturbative approaches and as a result some model functions for $\phi
_M(x,Q^2)$ may be proposed. Model distribution amplitudes for the mesons
are, therefore, objects of contradictory statements and conclusions made in
the literature. Such discussions is particularly intensive in the case of the
pion amplitude $\phi _\pi (x,Q^2)$ [10,11]. Nevertheless, in this article
for the pion we take ''traditional'' model distribution amplitudes proposed
in the literature [12-14]. They have the following form%
\begin{equation}
\phi _\pi (x,\mu _0^2)=\phi _{asy}^\pi (x)\left[
a+b(2x-1)^2+c(2x-1)^4\right] ,\label{3} 
\end{equation}
where $\phi _{asy}^\pi (x)$ is the pion asymptotic distribution amplitude,%
\begin{equation}
\phi _{asy}^\pi (x)=\sqrt{3}f_\pi x(1-x), \label{4}
\end{equation}
and $f_\pi $ is the pion decay constant; $f_\pi =0.093$ $GeV.$

The constants $a,b,c$ in Eq.(3) were found by means of the QCD sum rules
method at the normalization point $\mu _0=0.5$ $GeV$ and have the values
$$
a=0,~b=5,~c=0,~  CZ ~amplitude~ [12], 
$$
\begin{equation}
~~~~~~~ a=-0.1821,~b=5.91,~c=0,~ amplitude~ from~ Ref.[13], \label{5}
\end{equation}
$$
 a=0.6016,~b=-4.659,~c=15.52,~amplitude~ from~ Ref.[14]. 
$$

The meson distribution amplitude evolves in accordance with the expression
[12,15],%
\begin{equation}
\phi _M(x,Q^2)=\phi _{asy}(x)\sum^{\infty}_{n=0} 
r_nC_n^{3/2}(2x-1)\left[ \frac{\alpha _S\left( Q^2\right) }{\alpha _S\left(
\mu _0^2\right) }\right] ^{\gamma _n/\beta _0}. \label{6}
\end{equation}
In Eq.(6) $\left\{ C_n^{3/2}(2x-1)\right\} $ are the Gegenbauer polynomials, 
$\beta _0=11-2n_f/3$ is the QCD\ beta-function's first coefficient and $
\gamma _n$ is the anomalous dimension,
$$
\gamma _n=\frac 43\left[ 1-\frac 2{(n+1)(n+2)}+4\sum^{n+1}_{j=2}\frac 1j\right] . 
$$

The pion distribution amplitude (Eq.(3)) can be rewritten in the form (6).
But after defining of the coefficients $r_n$ and taking into account the
evolution of $\phi _\pi (x,Q^2)$ on $Q^2$, for our purposes it is
instructive to give to the distribution amplitude its old form, namely%
\begin{equation}
\phi _\pi (x,Q^2)=\phi _{asy}^\pi (x)\left[ \alpha +\beta (2x-1)^2+\gamma
(2x-1)^4\right] .\label{7} 
\end{equation}
In Eq.(7) the new coefficients $\alpha ,\beta ,\gamma $ are functions of $Q^2
$
$$
\alpha =a+\left[ \frac b5+\frac{14c}{105}\right] \left( 1-A_2\right) -\frac
c{21}\left( 1-A_4\right) , 
$$
\begin{equation}
~~~~~~~~~~~~~~~~~~~~~~~~\beta =bA_2+\frac{14c}{21}\left( A_2-A_4\right) , \label{8}
\end{equation}
$$
\gamma =cA_4. 
$$
Here $A_n$ is
$$
A_n=\left[ \frac{\alpha _S\left( Q^2\right) }{\alpha _S\left( \mu
_0^2\right) }\right] ^{\gamma _n/\beta _0} 
$$

Using the same method for the kaon we find%
\begin{equation}
\phi _K(x,Q^2)=\phi _{asy}^K(x)\left[ \alpha +\gamma \left( 2x-1\right)
+\beta (2x-1)^2+\delta (2x-1)^3\right] , \label{9}
\end{equation}
$$
\alpha =a+\frac b5\left( 1-A_2\right) ,\beta =bA_2,\gamma =\frac{3c}7\left(
A_1-A_3\right) ,\delta =cA_3. 
$$
In Eq.(9) $\phi _{asy}^K(x)$ has the same form as in Eq.(4), but with $f_\pi 
$ replaced by $f_K=0.122$ $GeV$ and constants $a,b,c$ take values [12]%
$$
a=0.4,~b=3,~c=1.25. 
$$

The QCD running coupling constant $\alpha_{S}(\hat{Q}^2)$ in Eq.(2) suffers
from ir singularities associated with the behaviour of the 
$\alpha_{S}(\hat{Q}^2)$ in the soft regions $x\to 0,~y \to 0;~~x \to 1,~y \to 1$.
Therefore, $F_{M}(Q^2)$ can be found after proper regularization of
$\alpha_{S}(\hat{Q}^2)$ in these soft end-point regions. For these purposes
let us relate the running coupling $\alpha_{S}({\lambda}Q^2)$ in terms of
$\alpha_{S}(Q^2)$ by means of the renormalization group equation
\begin{equation}
\alpha _S(\lambda Q^2)\simeq \frac{\alpha _S(Q^2)}{1+\left( \alpha
_S(Q^2)\beta _0/4\pi \right) \ln \lambda }. \label{10}
\end{equation}
where $\alpha_{S}(Q^2)$ is the one-loop QCD coupling constant.

{\bf 3.} As was demonstrated in our previous works [6-8], integration in Eq.(1)
using Eqs.(2),(10) generates ir divergences and as a result for $F_{M}(Q^2)$
we get a perturbative series with factorially growing coefficients. This
series can be resummed using the Borel transformation [2], 
\begin{equation}
\left[ Q^2F_M(Q^2)\right] ^{res}=\frac{(16\pi f_M)^2}{\beta _0}\int_{0}^{\infty} du\exp
\left( -\frac{4\pi u}{\beta _0\alpha _S(Q^2)}\right) B\left[ Q^2F_M\right]
(u), \label{11}
\end{equation}
where $B[Q^2F_{M}](u)$ is the Borel transform of the corresponding perturbative 
series [7,8].

But simple trick allow us from Eq.(1) directly obtain Eq.(11). In fact, after 
changing in Eq.(1) the variables $x$ and $y$ to $z=\ln (1-x)$ an $w=\ln (1-y)$
(or $z=\ln x,~w=\ln y$) and using formula
$$
\int_{0}^{\infty} e^{-u\left( t+w+z\right) }du=\frac 1{t+w+z},~~t=\frac{4\pi }{\beta
_0\alpha _S(Q^2)}, 
$$
and after integration over $z,~w$ we find for the meson elm form factor the expression
(11), where the Borel transform has the form
\begin{equation}
B\left[ Q^2F_M\right] (u)=\sum^N_{n=1}\left( \frac{%
{\bf m}_n(Q^2)}{(n-u)^2}+\frac{{\bf l}_n(Q^2)}{n-u}\right) . \label{12}
\end{equation}
The exact expressions for ${\bf m}_n(Q^2), {\bf l}_n(Q^2)$ are given
in the Appendix.

The Borel transform (12) has double and single poles at $u=n$. Then the resummed
expression (11) can be calculated with the help of the principal value prescription [3],
\begin{equation}
\left[ Q^2F_M(Q^2)\right] ^{res}=\frac{(16\pi f_M)^2}{\beta _0}\sum^{N}_{n=1}
\left[ -\frac{{\bf m}_n}n+({\bf l}_n+{\bf m}_n\ln
\lambda )\frac{Li(\lambda ^n)}{\lambda ^n}\right] , \label{13}
\end{equation}
where $Li(\lambda)$ is the logarithmic integral [16],
\begin{equation}
Li(\lambda )=P.V.\int_{0}^{\lambda} \frac{dx}{\ln x},~~\lambda =Q^2/\Lambda ^2. \label{14}
\end{equation}

In Eq.(13) we have taken into account the dependence of the distribution 
amplitude $\phi_M(x,Q^2)$ on the scale $Q^2$ and it is the generalization of
our results for the pion and kaon elm form factor. Indeed, if we switch off this
dependence ($A_n \equiv 1$), then Eq.(13) coincides for $a=0,~b=5,~c=0$ and $N=4$ with
our result for the pion Ref.[7] and for $a=0.4,~b=3,~c=1.25$ and $N=5$ with the
kaon elm form factor obtained in Ref.[8].

{\bf 4.} For studing phenomenological consequences of the ir renormalon
contributions to $Q^2F_M(Q^2)$, it is instructive to introduce the ratio
$R_M=[Q^2F_M(Q^2)]^{res}/$\\$[Q^2F_M(Q^2)]^0$, where $[Q^2F_M(Q^2)]^0$ is the meson
elm form factor calculated in the context of the frozen coupling approximation.

In the frozen coupling approximation, for example, the pion form
factor is 
\begin{equation}
Q^2F_\pi =64\pi f_\pi ^2\left\{ \frac a2+\frac b{10}+\frac{3c}{70}+\frac
1{15}\left( b+\frac{2c}3\right) A_2+\frac{4c}{15\cdot 21}A_4\right\} ^2. \label{15}
\end{equation}

In Fig.1 the dependence of the ratio $R_\pi$ on $Q^2$ (CZ distribution amplitude)
is shown. As is seen the
ir renormalon contributions enhance the perturbative result and are 
considerable for all values of $Q^2$, where the running coupling constant 
method is applicable $(Q^2/ \Lambda^2>>1, ~Q^2 \geq 2~GeV^2)$. The dependence
of the distribution amplitude on the scale $Q^2$ changes only the numerical
results for the ratio $R_\pi$, particularly in the region $Q^2 \geq 5~GeV^2$,
preserving at the same time the shape of the curve.
The ir corrections can be transferred into the scale of 
$\alpha_{S}(Q^2)$ in Eq.(15) 
$$
Q^2\rightarrow e^{f(Q^2)}Q^2,~f(Q^2)=c_1+c_2\alpha _S(Q^2). 
$$
Numerical fitting allow us to get (for $n_f=3$) the values of $c_1,~c_2$;
$$c_1 \simeq -8.528,~c_2 \simeq 26.28$$.

The similar numerical results can be obtained also for the $R_\pi$ using other
pion distribution amplitudes.
For the kaon a more reliable form of $f(Q^2)$ is [8]
$$f(Q^2)=c_1+c_2\alpha_S(Q^2)+c_3\alpha^2_S(Q^2)$$
where the coefficients found after numerical fittings are
$$c_1 \simeq -1.304,~c_2 \simeq -35.604,~ c_3 \simeq 127.25$$
The results of numerical calculation for the kaon are depicted in Fig.2.

The obtained results have important consequences for a fate of the hard-
scattering model based on pQCD at intermediate (moderate) energies 
($2~GeV^2\leq Q^2 \leq 10~GeV^2$). It is well
known that Eq.(1) was derived in the framework of pQCD by neglecting transverse
momenta of meson's constituents and higher Fock components of the distribution
amplitude [15]. In Ref.[17] a modified perturbative QCD approach was proposed,
which takes into account the partonic transverse momenta, as well as the Sudakov
corrections. In the context of this approach the pion electromagnetic form
factor was reexamined in works [18,19]. In these papers the authors used model 
light-cone wave functions and included also the unconventional helicity
components $h_1+h_2=\pm 1$ of the pion wave function (see, Ref.[19]). The authors
of Ref.[18] made conclusion that the perturbative predictions are smaller than the 
experimental data, or that adding the $h_1+h_2=\pm 1$ components suppresses 
the hard-scattering significantly and non-perturbative contributions will
dominate in the present experimentally-accessible energy region [19]. 
But in these works in calculations the frozen coupling approximation was
applied. We think that the running coupling constant method may improve the 
situation with the modified pQCD in the region
of moderate $Q^2$, because the ir renormalon effects work in opposite
direction than factors considered in Refs.[18,19]. Therefore, we do not agree
with Ref.[19], where the authors rule out the hard-scattering model for describing
the light mesons form factors at moderate $Q^2$.

Our results do not allow us also to exclude the CZ distribution amplitude as the
reliable model function, because in deriving of Eqs.(12),(13) we have neglected
the second term in Eq.(10) (Contopanagos, Sterman, Ref.[3]), as well as the second
order corrections to $T_H$ [20]; before comparison with the experimental data [21], 
these points must be clarified.
\newpage
{\Large \bf APPENDIX}

The functions ${\bf m}_n(Q^2)$, ${\bf l}_n(Q^2)$ for the pion have the forms
(N=6),%
$$
{\bf m}_1(Q^2)=\left( \alpha +\beta +\gamma \right) ^2,{\bf m}_2(Q^2)=\left(
\alpha +5\beta +9\gamma \right) ^2,{\bf m}_3(Q^2)=64\left( \beta +9\gamma
\right) ^2, 
$$
$$
{\bf m}_4(Q^2)=16\left( \beta +14\gamma \right) ^2,{\bf m}_5(Q^2)=2304\gamma
^2,{\bf m}_6(Q^2)=256\gamma ^2, 
$$
and%
$$
{\bf l}_1(Q^2)=-2\left( \alpha +\beta \right) \left( \alpha +\frac 73\beta
\right) -\frac 2{15}\gamma \left( 58\alpha +90\beta +43\gamma \right) , 
$$
$$
{\bf l}_2(Q^2)=2\left( \alpha ^2-25\beta ^2\right) -2\gamma \left( 6\alpha
+120\beta +135\gamma \right) , 
$$
$$
{\bf l}_3(Q^2)=8\beta \left( \alpha +\beta \right) +\frac 83\gamma \left(
12\alpha -161\beta -692\gamma \right) , 
$$
$$
{\bf l}_4(Q^2)=\frac 43\beta \left( 35\beta -\alpha \right) -\frac 43\gamma
\left( 14\alpha -417\beta +1022\gamma \right) , 
$$
$$
{\bf l}_5(Q^2)=8\gamma \left( \alpha +17\beta +321\gamma \right),~{\bf l_6}%
(Q^2)=\frac 8{15}\gamma \left( 1717\gamma -3\alpha -20\beta \right) . 
$$

For the kaon we get (N=5),%
$$
{\bf m}_1(Q^2)=\left( \alpha +\beta \right) ^2+\left( \gamma +\delta \right)
^2+\frac 23\left( \alpha +\beta \right) \left( \gamma +\delta \right) , 
$$
$$
{\bf m}_2(Q^2)=\left( \alpha +5\beta \right) ^2+\left( 3\gamma +7\delta
\right) ^2+\frac 23\left( \alpha +5\beta \right) \left( 3\gamma +7\delta
\right) , 
$$
$$
{\bf m}_3(Q^2)=64\beta ^2+4\left( \gamma +9\delta \right) ^2+\frac{32}3\beta
\left( \gamma +9\delta \right) , 
$$
$$
{\bf m}_4(Q^2)=16\beta ^2+400\delta ^2+\frac{160}3\beta \delta ,~{\bf m}_5
(Q^2)=64\delta ^2, 
$$
and
$$
{\bf l}_1(Q^2)=-2\left( \alpha +\beta \right) \left( \alpha +\frac 73\beta
\right) -4\left( \gamma +\delta \right) \left( \gamma +\frac 43\delta
\right) -\frac 29\left( 9\alpha \gamma +11\alpha \delta +13\beta \gamma
+15\beta \delta \right) , 
$$
$$
{\bf l}_2(Q^2)=2\left( \alpha ^2-25\beta ^2\right) -2\left( 3\gamma +7\delta
\right) \left( \gamma +\frac{29}3\delta \right) +\frac 29\left( 6\alpha
\gamma -8\alpha \delta -60\beta \gamma -250\beta \delta \right) , 
$$
$$
{\bf l}_3(Q^2)=8\beta \left( \alpha +\beta \right) +2\left( \gamma +9\delta
\right) \left( 5\gamma -19\delta \right) +\frac 23\left( \alpha \gamma
+9\alpha \delta +21\beta \gamma -67\beta \delta \right) , 
$$
$$
{\bf l}_4(Q^2)=\frac 43\beta \left( 35\beta -\alpha \right) +\frac{20}%
3\delta\left( 5\gamma +41\delta \right) +\frac 29\left( -10\alpha \delta +10\beta
\gamma +432\beta \delta \right) , 
$$
$$
{\bf l}_5(Q^2)=4\delta \left( -\gamma +\frac{157}3\delta \right) +\frac
49\delta \left( \alpha +17\beta \right) . 
$$

\newpage
{\Large \bf REFERENCES}\vspace{10mm}\\
{\bf 1.} M.Neubert, Phys.Rev.D51 5924 (1995);\\
P.Ball, M.Beneke and V.M.Braun, Nucl.Phys.B452 563 (1995); Phys.Rev.D52 3929 (1995);\\
M.Beneke and V.M.Braun, Phys.Lett.B348 513 (1995);\\
C.N.Lovett-Turner and C.J.Maxwell, Nucl.Phys.B432 147 (1994); B452 188 (1995);\\
G.P.Korchemsky and G.Sterman, Nuc.Phys.B437 415 (1995);\\
B.R.Webber, Renormalon Phenomena in Jets and Hard Processes, Talk given at 27th ISMD97, hep-ph/9712236;\\
V.I.Zakharov, Renormalons as a Bridge between Perturbative and Nonperturbative 
Physics, Talk presented at YKIS97, Kyoto, 97, hep-ph/9802416.\\
{\bf 2.} G. 't Hooft, In. "The Whys of Subnuclear Physics", Proc. Int. School,
Erice, 1977, ed. A.Zichichi, Plenum, New York, 1978;\\
V.I.Zakharov, Nucl.Phys.B385 452 (1992).\\
{\bf 3.} A.H.Mueller, Nucl.Phys.B250 327 (1985);\\
H.Contopanagos and G.Sterman, Nucl.Phys.B419 77 (1994).\\
{\bf 4.} S.J.Brodsky, G.P.Lepage and P.B.Mackenzie, Phys.Rev.D28 228 (1983).\\
{\bf 5.} B.R.Webber, Phys.Lett.B339 148 (1994);\\
Yu.L.Dokshitzer and B.R.Webber, Phys.Lett.B352 451 (1995);\\
Yu.L.Dokshitzer, G.Marchesini and B.R.Webber, Nucl.Phys.B469 93 (1996);\\
M.Dasgupta, B.R.Webber, Nucl.Phys.B484 247 (1997). \\
{\bf 6.} S.S.Agaev, Phys.Lett.B360 117 (1995); E.Phys.Lett.B369 379 (1996);\\
S.S.Agaev, Mod.Phys.Lett.A10 2009 (1995).\\
{\bf 7.} S.S.Agaev, ICTP preprint IC/95/291, 1995, hep-ph/9611215 (unpublished).\\
{\bf 8.} S.S.Agaev, Mod.Phys.Lett.A11 957 (1996).\\
{\bf 9.} S.S.Agaev, Eur.Phys.J.C1 321 (1998).\\
{\bf 10.} V.M.Belyaev and M.B.Johnson, SPhT preprint, SPhT-t97, hep-ph/9707329;\\
          A.V.Radyushkin, Report JLAB-THY-97-29, hep-ph/9707335.\\
{\bf 11.} S.J.Brodsky, C.-R.Ji, A.Pang and D.G.Robertson, SLAC preprint, SLAC-PUB-7473, hep-ph/9705221.\\
{\bf 12.} V.L.Chernyak and A.R.Zhitnitsky, Phys.Rep.112 173 (1984).\\
{\bf 13.} G.R.Farrar, K.Huleihel and H.Zhang, Nucl.Phys.B349 655 (1991).\\
{\bf 14.} V.M.Braun and I.E.Filyanov, Z.Phys.C44 157 (1989).\\
{\bf 15.} G.P.Lepage and S.J.Brodsky, Phys.Rev.D22 2157 (1980);\\
A.Duncan and A.H.Mueller, Phys.Rev.D21 1626 (1980);\\
A.V.Efremov and A.V.Radyushkin, Phys.Lett.B94 245 (1980).\\
{\bf 16.} A.Erdelyi et al., Higher Transcendental Functions, (McGraw-Hill, 1953), Vol.2\\
{\bf 17.} H.N.Li and G.Sterman, Nucl.Phys.B381 129 (1992).\\
{\bf 18.} F.-G.Cao, T.Huang and C.-W.Luo, Phys.Rev.D52 5358 (1995).\\
{\bf 19.} S.W.Wang and L.S.Kisslinger, Phys.Rev.D54 5890 (1996).\\
{\bf 20.} R.D.Field, R.Gupta, S.Otto and L.Chang, Nucl.Phys.B186 429 (1981).\\
{\bf 21.} L.J.Bebek et al., Phys.Rev.D9 1229 (1974); D13 25 (1976);
D17 1693 (1978).\\
\newpage
{\Large \bf FIGURE CAPTION}\vspace{10mm}\\
{\bf Fig.1} The ratio $R_{\pi}$ for the pion (CZ amplitude) is shown as a 
function of $Q^2$. 
The solid curves {\bf 1} and {\bf 2} correspond to $R_{\pi}$ with 
$[Q^2F_{\pi}(Q^2)]^0$ in
the frozen coupling approximation and after the scale-setting procedure,
respectively. The dashed curve is the ratio $R_{\pi}$ from Ref.[7], where the
dependence of the pion distribution amplitude on the hard scale $Q^2$ was 
neglected. In calculations the QCD parameter $\Lambda$ is taken to be 
equal to $0.1~GeV$.\\
{\bf Fig.2} The same, but for the kaon. The dashed curve is taken from
Ref.[8].

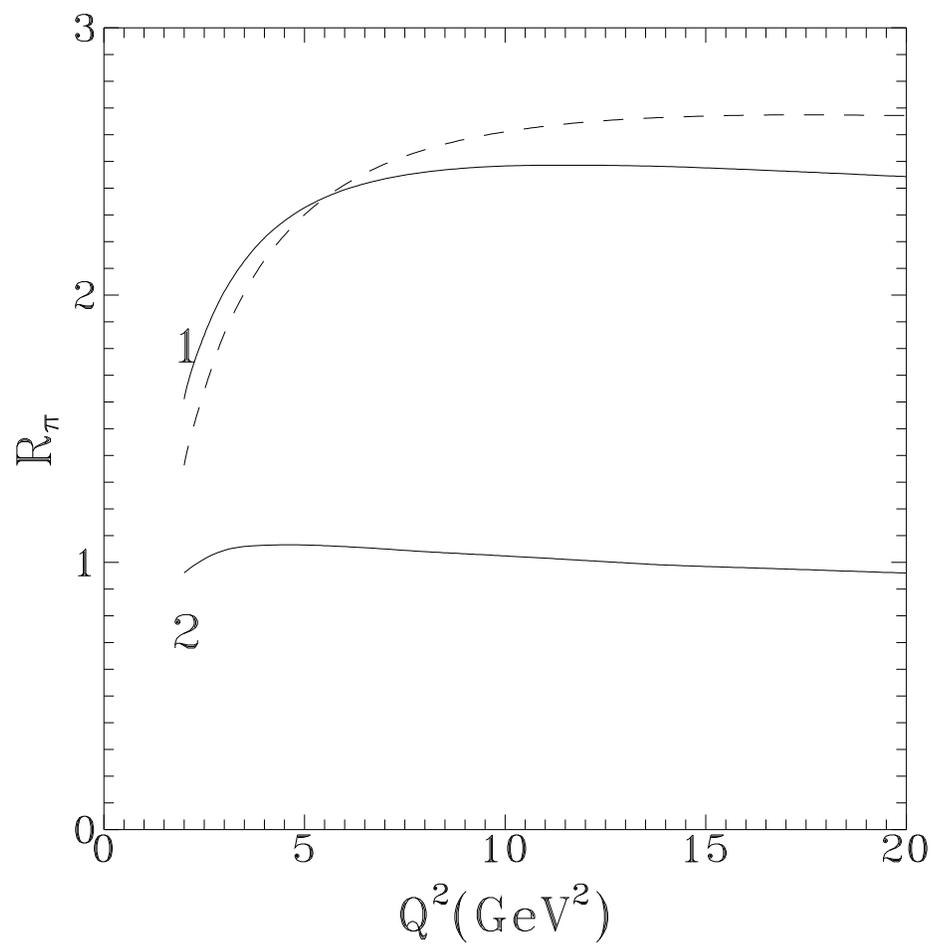
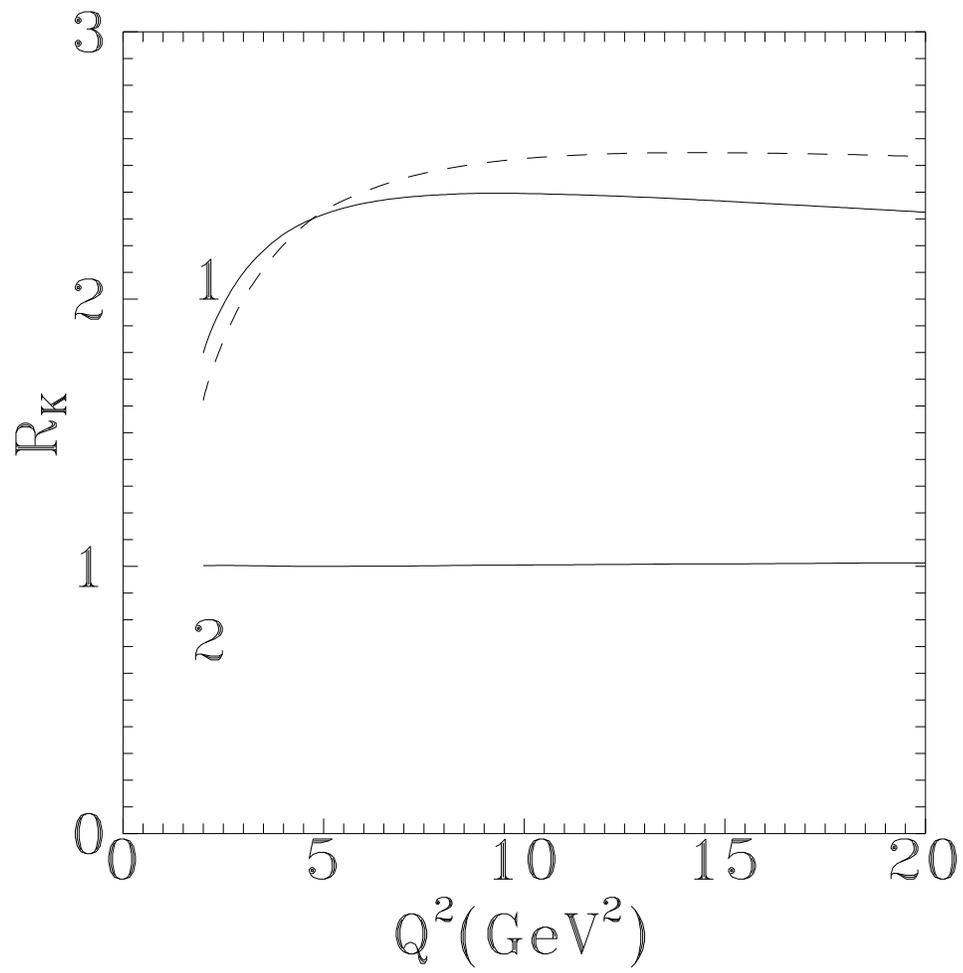
\end{document}